%% file: main.tex
\documentclass[10pt,letterpaper]{article}

\usepackage{titlesec}
\usepackage{url}
\usepackage[usenames,dvipsnames]{color}
\usepackage{authblk}
\usepackage{pdflscape}
\usepackage[table,xcdraw]{xcolor}
\usepackage[justification=centering]{caption}
\usepackage{graphicx}
\usepackage{wrapfig}
\usepackage{diagbox}
\usepackage{tabularx}
\usepackage{float}
\usepackage[skins,breakable]{tcolorbox}
\usepackage{todonotes}
\usepackage{soul,color}

\usepackage{listings} 
\definecolor{codegreen}{rgb}{0,0.6,0}
\definecolor{codegray}{rgb}{0.5,0.5,0.5}
\definecolor{codepurple}{rgb}{0.58,0,0.82}
\definecolor{backcolour}{rgb}{0.96,0.96,0.96}
\definecolor{linecolour}{rgb}{0.9,0.9,0.9}
\lstdefinestyle{mystyle}{
	language=R,                     
	basicstyle=\small\ttfamily, 
	numberstyle=\tiny\color{blue},  
	stepnumber=1,                   
	numbersep=5pt,                  
	backgroundcolor=\color{backcolour},  
	showspaces=false,               
	showstringspaces=false,         
	showtabs=false,                 
	frame=single,                   
	rulecolor=\color{linecolour},        
	tabsize=2,                      
	captionpos=b,                   
	breaklines=true,                
	breakatwhitespace=false,        
	keywordstyle=\color{blue},      
	commentstyle=\color{codegreen},   
	stringstyle=\color{red},      
	morekeywords={head,tail},
	deletekeywords={model,par,coefficients,c,nrow,sum,split,set,as,integer,exp,log,length,data,frame}
} 
\usepackage[normalem]{ulem} 

\usepackage{geometry} 

\usepackage{amsmath,amssymb}

\usepackage{changepage}

\usepackage[utf8x]{inputenc}

\usepackage{textcomp,marvosym}

\usepackage{cite}

\usepackage{nameref,hyperref}

\usepackage[right]{lineno}

\usepackage{microtype}
\DisableLigatures[f]{encoding = *, family = * }

\usepackage{array}

\usepackage{setspace} 

\usepackage[aboveskip=1pt,labelfont=bf,labelsep=period,justification=raggedright,singlelinecheck=off]{caption}

\def\tcb@proc@counter@auto#1{%
  \newcounter{tcb@cnt@#1}%
  \csxdef{tcb@cnt@#1}{tcb@cnt@#1}%
  \tcb@proc@counter@autoanduse{#1}%
  \ifcsname resetcounteronoverlays\endcsname
  \resetcounteronoverlays{tcb@cnt@#1}
  \fi
}
\makeatother
\newtcolorbox[auto counter]{numberedbox}[2][]{%
breakable,
title=Box~\thetcbcounter: #2,#1}

\title{Nine tips for ecologists using machine learning}
\author[1]{Marine Desprez}
\author[2,*]{Vincent Miele}
\author[1]{Olivier Gimenez}

\affil[1]{CEFE, Univ Montpellier, CNRS, EPHE, IRD, Montpellier, France}
\affil[2]{Universit\'e de Lyon, F-69000 Lyon; Universit\'e Lyon 1; CNRS, UMR5558, Laboratoire de Biom\'etrie et Biologie \'Evolutive, F-69622 Villeurbanne, France}
\affil[*]{Corresponding author: Vincent.Miele@univ-lyon1.fr}

\date{}

\titleformat{\section}
  {\normalfont\Large\bfseries}{Tip \thesection:}
  {1em}{}

\begin{document}
\maketitle

\begin{abstract} 
Due to their high predictive performance and flexibility, machine learning models are an appropriate and efficient tool for ecologists. However, implementing a machine learning model is not yet a trivial task and may seem intimidating to ecologists with no previous experience in this area. Here we provide a series of tips to help ecologists in implementing machine learning models. We focus on classification problems as many ecological studies aim to assign data into predefined classes such as ecological states or biological entities. Each of the nine tips identifies a common error, trap or challenge in developing machine learning models and provides recommendations to facilitate their use in ecological studies.
\end{abstract}

\section*{Introduction}\label{intro}

Ecological datasets are generally characterised by complex interactions between variables, non-linearity, missing values, dependence in the observations and/or a continuously expanding size \cite{cutler2007random, roberts2017cross, christin2019applications}, especially since the recent increase in the use of remote sensing and automatic recorders \cite{rovero2013camera}. A growing number of those datasets cannot be effectively processed by humans anymore and require methods that can deal with high number of variables and complex data structures \cite{schneider2020three, christin2019applications, geirhos2020shortcut}. Because of their ability to process large and complicated datasets, machine learning models are expected to become a standard framework in the analysis of ecological data \cite{christin2019applications, humphries2018machine,pichler2022machine}. 
Over the last few years, machine learning algorithms have become increasingly popular due to their high performance and flexibility \cite{pichler2022machine}. In ecology, they have been successfully applied to perform various tasks such as identifying species from images or sounds \cite{waldchen2018machine}, monitoring animal behaviour \cite{valletta2017applications} or modelling species distribution \cite{gobeyn2019evolutionary} and new innovative studies and perspectives keep being regularly documented \cite{christin2019applications,tuia2022perspectives}.

However, implementing a machine learning model is not yet a trivial task and may seem intimidating to ecologists with no previous experience in this area. In this paper, we aim to share nine tips to help ecologists avoid some of the most common errors and incorrect practices in machine learning.
We focused our tips on 
classification problems as a substantial number of ecological studies aim to assign data into predefined classes such as ecological states or biological entities. Some typical examples of classification include species identification through pictures \cite{waldchen2018machine} or sound recordings \cite{dugan2015dcl, shamir2014classification, acevedo2009automated}, distinction of different phenological phases in plant life cycle \cite{czernecki2018machine, xin2020evaluations}, description of animal behaviour \cite{norouzzadeh2018automatically} and detection of disease in plants \cite{mohanty2016using}. 
Each tip presented in this paper identifies a common error or challenge in developing machine learning models and provides recommendations to facilitate the use of machine learning methods in ecological studies.\\

\section{Adopt the machine learning mindset}\label{MLmindset}

The concept behind machine learning refers to the use of a certain type of model that can discover and {\it learn} patterns in data to generate predictions or detect patterns automatically without having to follow explicit instructions. Without machine learning, humans have to provide data and instructions; with machine learning, humans have to provide data.

The learning phase can happen in two different ways, with or without supervision  (see \cite{pichler2022machine} for an introductory review). In unsupervised learning, the model automatically discovers patterns and similarities in unlabelled data (e.g. data for which we do not have a label indicating the associated class, given by the user). Unsupervised learning is often used in data exploration to find the underlying structure of the dataset, reduce its dimensions or cluster/group similar data together. The state-of-the-art methods include PCA, k-means and hierarchical clustering, or the more recent methods  t-SNE \cite{van08} and UMAP\cite{mci18} that are particularily popular in this context. 
The latter was for instance used to compare soundscapes from a variety of ecosystems \cite{set20}. In supervised learning, a labelled dataset is initially provided to the model: those labelled data include input variables and output variables (e.g. class labels). These data play the role of a supervisor that teaches the model how to correctly predict the output by finding a function that maps the explanatory input variables with the output. Depending on whether the output is a discrete category or a quantity, the problem is called a classification or a regression problem respectively.  Random forests, XGBoost and neural networks are leading options in this framework. In ecology, a classical example of a supervised learning task is the classification of individuals in different categories based on a set of explanatory variables. This problem can be solved using a wide range of models from logistic models to complex deep learning models.

The machine learning mindset can be presented by revisiting a classification problem (see a concrete example in Box \ref{logisticbox}). Before being able to predict to which category a new individual belongs, the model (logistic regression in Box \ref{logisticbox}) has to enter a learning phase that consists in finding the optimal parameters describing the relationship between the variables and the labelled output. How does the model find those optimal parameters? By minimizing a loss function. This function (e.g. mean squared error function), the keystone of a supervised learning approach, evaluates how far from the correct answers/outputs are the model predictions. The learning phase consists in minimizing this loss function numerically, such that the optimal parameters (i.e., those that lead to the lowest predictive error) are the ones chosen for the final model. This final model can then be used to predict the labels of a new set of individuals.\\

\section{Create your data sets (very) carefully}\label{splitDataset}
\begin{figure}[H]
\centering
\includegraphics[scale = 0.99]
{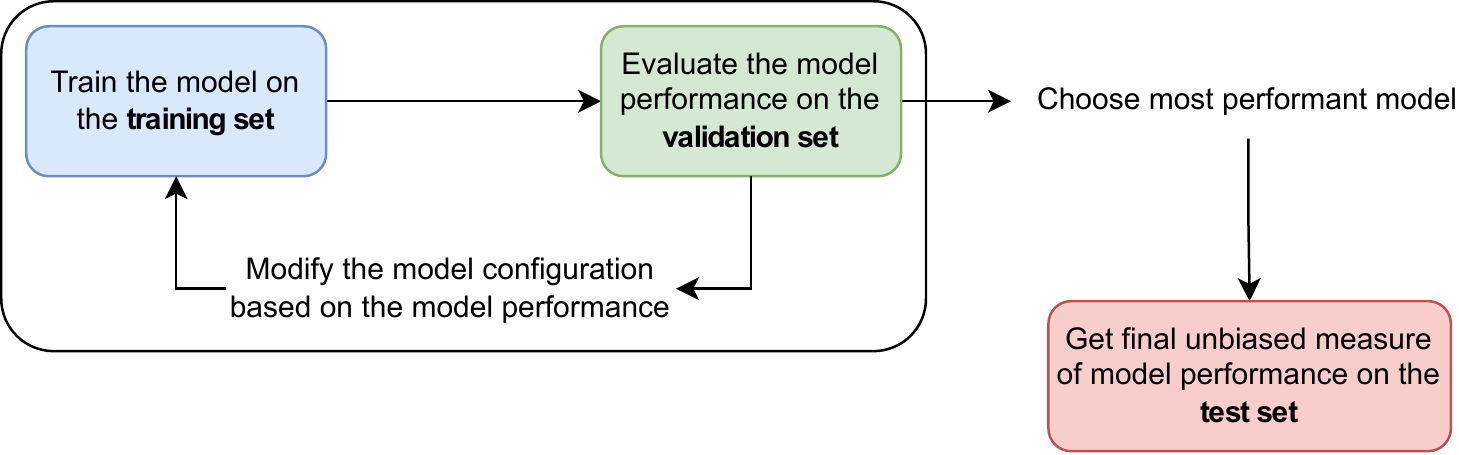}\vspace{0.3cm}
\caption{Illustration of the different steps of developing a machine learning model involving three separate sets of data.} 
\label{fig_model_building}
\end{figure}

Here, we focus on classification tasks but the following principles are also applicable to regression problems.
The general approach to solve a classification problem in machine learning is to: (i) develop different versions of a method of classification (a classifier) and \textit{train} them on a dataset; (ii) evaluate and compare the models’ predictive performance using an evaluation metric (see Tip \ref{evalMetric}) and (iii) select the best performing model to carry out the final predictions on a sample of new unseen dataset (Fig. \ref{fig_model_building}). Before entering the learning process, the collected data should be composed of three separate sets: the training set, the validation set and the test set. This  partitioning is necessary because the data used to train and evaluate the model need to remain independent in order to obtain reliable performance measure \cite{gareth2013introduction, chr21}.

\paragraph{The three separate data sets.}
{\bf Training.} Only the training data set is used to train the model (Fig. \ref{fig_model_building}). Ideally, this training set should include a various set of inputs to train the model under as many situations as possible in order to predict any unseen data sample that may appear in the future. 
{\bf Validation.} The trained model is simultaneously used to predict the classes from the observations in the separate validation set to evaluate the model predictive performance (Fig. \ref{fig_model_building}). Evaluating the model on a separate validation set prevents the model from overfitting, i.e., when the model memorizes the pattern in the training data to such an extent that it fails to generalize and to make accurate predictions on unseen data \cite{ying2019overview, dietterich1995overfitting}. As an example, imagine a model developed to detect the presence of dogs in pictures. The model performance obtained on the training set is high, indicating the model is doing a good job classifying the pictures with and without dogs. However, when tested on the separate validation set, its performance drops. This indicates that the model was overfitting; it memorized specific patterns of the dogs from the training set instead of learning general patterns common to dogs. 
This issue would not have been detected if the model was evaluated on the same data it was trained on. During the training (or tuning) phase, different models or set of variables may be tested and hyperparameters optimised (e.g., the number of trees in a random forest or the number of layers in a neural network). The simultaneous validation phases provide information about how those different model configurations affect the model predictive performance. Training and validation phases are repeated until a desirable predictive performance is reached on the validation set. 
{\bf Test.} In a final step, the best performing model is run on the test set to obtain an unbiased measure of the model predictive performance (Fig. \ref{fig_model_building}). It is critical that the test set (sometimes called out-of-sample, \cite{whytock2021robust}) remains out of the development phase until the final predictions are made for the performance measure to be reliable \cite{kuhn2013applied,kapoor2022leakage,chr21}.

\paragraph{The good, the bad and the ugly data sets ?} 
The validation set is used to confirm the training process is correct, and that the model choice is satisfying. The main issue with this validation set is to avoid data leakage (see Tip \ref{dataLeakage}) that can lead to over-optimistic model evaluation. However, it should not be used to claim that the trained model will perform well when deployed in real scenarios. This is the role of the test dataset. Test sets must be representative of the target data, i.e. data in real life for which the model was built. The key recommendation to obtain a reliable measure of model performance, is to keep the test data set as independent as possible. For example, using camera trap data from different locations independently, or continuous sections of longitudinal data from dates beyond the end of the training set  
\cite{roberts2017cross}. A random splitting of dataset into train/validation/test sets is therefore strongly discouraged.
 
If the training data is an unbiased sample of the underlying distribution, then the learned classification function will generalize well and will make accurate predictions for new samples \cite{pichler2022machine,gareth2013introduction, ying2019overview}. If not, the distribution of the target data may differ from the distribution of the training data and the classification function will perform poorly \cite{kouw2018introduction}.
This issue, called distribution or domain shift \cite{machireddy2022continual}, is a common cause of a well-known scenario in machine learning: a seemingly impressive model (as evaluated on validation data) that completely fails when used on a test set of new data \cite{pichler2022machine}. 
In species identification for example, a model trained on pictures with clean weather conditions will likely fail on a test set of pictures with adverse weather conditions (e.g., rain, fog, snow \cite{machireddy2022continual}).  The solution often consists in enlarging the training set with new data covering the complete distribution expected for the data in real conditions.

Choosing appropriate validation and test sets is one of the most important step in a machine learning project. We cannot stress this enough. A poor choice of validation and test sets will lead to a disconnect between the results in development and the ones obtained when deploying the model on new data \cite{chr21}.


\section{Get the right amount of data}\label{quantityData} 
The amount of data needed to capture the relationship between the explanatory variables and the output data varies depending on the complexity of the problem and model. While there is no general rule for determining the quantity of data required in machine learning models, three concepts should be kept in mind when gathering data for the training, validation and test steps. 

First, machine learning models, especially deep learning models \cite{lecun2015deep}, often need a significant amount of training data. This is because the number of parameters in those models can be tremendous (tens of millions for most convolutional neural networks (CNN)). Therefore, complex models will need significantly more data than simpler ones (e.g. thousands or millions of pictures  to train a model for species identification \cite{rigoudy2022deepfaune, norouzzadeh2018automatically,waldchen2018machine}). 
Second, a model can only capture what it is trained on. For instance, a model trained on daytime images will not work on night images; a species distribution model trained on mountains will not work on wetlands. Therefore, the training set should include as much diversity/variability and edge cases as possible to enable the model to learn and predict various scenarios. Note that a large dataset does not necessarily include sufficient variability in the data to guarantee good model performance \cite{volk2019towards}.
For instance, a dataset might contain millions of pictures of a given species, but only in sunny conditions. Using this dataset to identify that species during rainy or snowy days would likely produce poor predictions (see Tip \ref{shortcutlearning}). 
Therefore, we recommend to  pay a greater attention to the variability/diversity available, not to only focus on the raw amount of data. 
Third, if there is not enough data to build correct validation and test sets, the model evaluation metric will have a greater variance which will (i) prevent a proper tuning of the model and (ii) make it hard to assess how well the model will perform on new data and generalize. 

If more data is needed, we suggest to make use of existing methods to increase the sample size. This includes data augmentation techniques to generate heterogeneous data from existing training data \cite{mumuni2022data}, crowd sourcing to maximize data sources \cite{uhlmann2019scientific} or creation of large consortium to gather more data \cite{rigoudy2022deepfaune}). It can also consist in going back to the field to collect more labelled data (when possible). 

In deep learning specifically, an alternative approach is to limit the amount of training data required by (re-)using existing pre-trained models as a starting point for new models using a transfer learning approach \cite{tod23}. Recently developed self-supervised learning methods \cite{pan21} are also another option to solve the challenges posed by the needs of large labelled data. They consists in learning from the similarity between close images (e.g. two sub-parts of the same image, two successive camera trap images \cite{pan21}) and can cope with the limited availability of some categories (e.g. rare species).

\section{Be mindful of data leakage}\label{dataLeakage}
Data leakage is one of the leading machine learning errors (\cite{kapoor2022leakage}). It happens when a model is trained using data that contain some information  that would not be available at the time of predictions \cite{kaufman2012leakage}. It is a serious problem as it can create overly optimistic, if not completely invalid, predictive models with very poor generalization.

\begin{figure}[H]
\centering
\includegraphics[scale=0.8]
{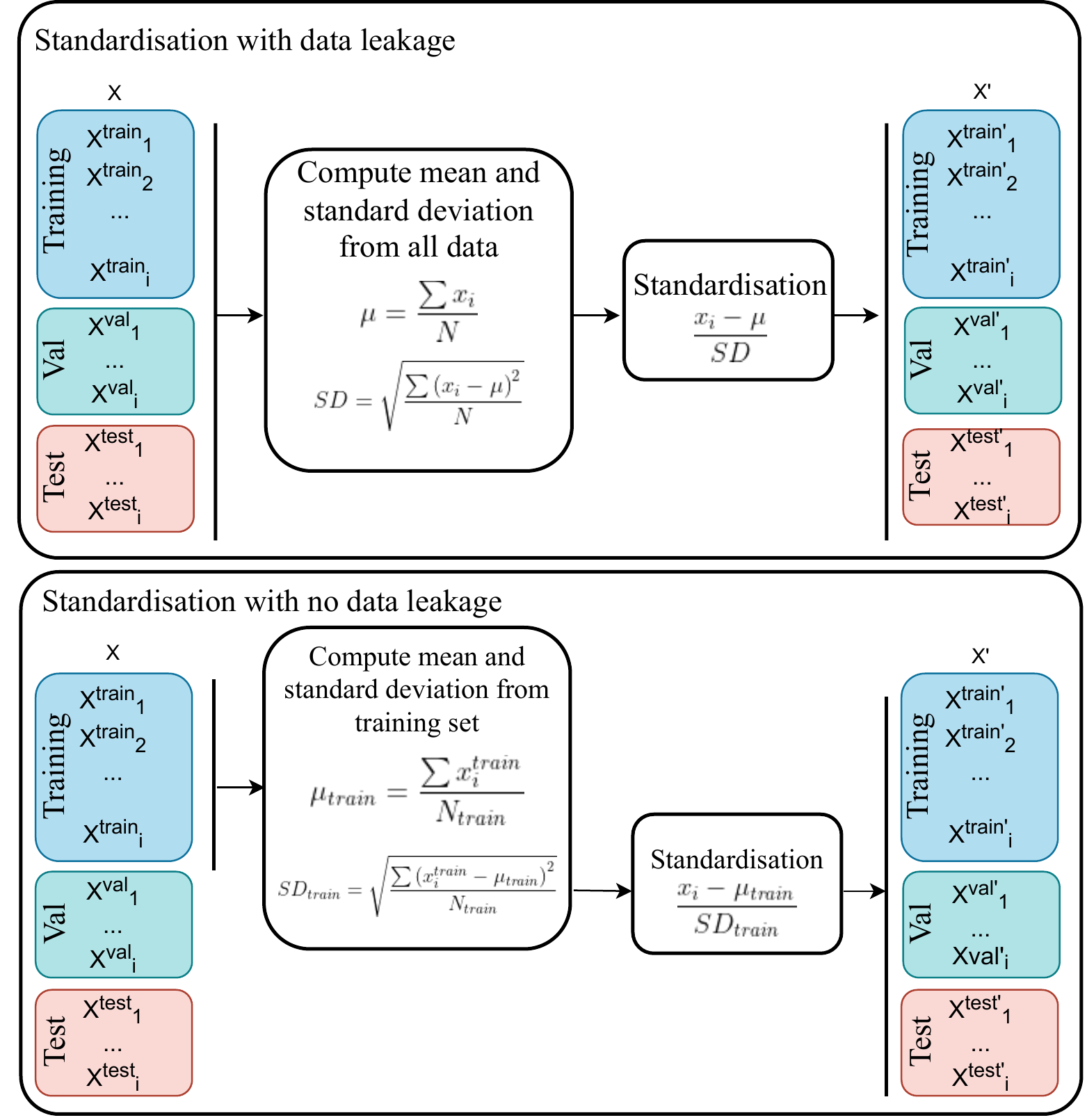}\vspace{0.3cm}
\caption{Data standardisation computed with and without data leakage}
\label{fig_data_leak_ex_std}
\end{figure}

\begin{figure}[H]
\centering
\includegraphics[scale=0.5]
{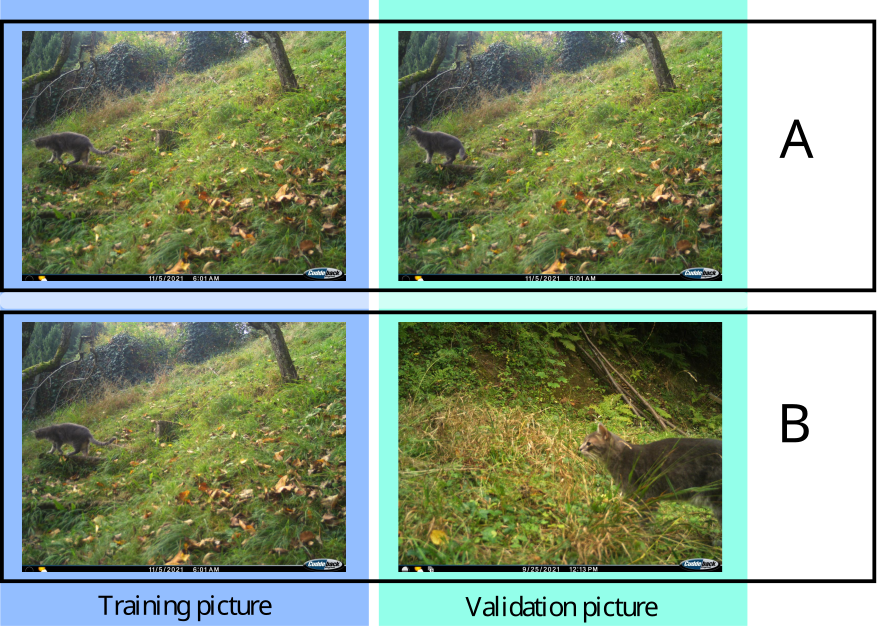}\vspace{0.3cm}
\caption{
Example of data leakage due to data duplication. In case A, pictures in training and validation sets are from the same temporal sequence and look near identical.
In that case, the model will work artifactually well on the validation set. In case B however, the validation set is informative since training and validation sets are independent (no duplication). Source: Vincent Miele}
\label{fig_data_leak_duplication}
\end{figure}

In practice, data leakage often occurs subtly and inadvertently, making it hard to detect and eliminate \cite{kaufman2012leakage}. A common example of data leakage is when the pre-processing of data is done on the whole combination of training, validation and test sets. Knowledge of the full data distribution is included in the processed data and used by the model (see Fig. \ref{fig_data_leak_ex_std} for an example on standardisation). A non-leaky way of processing the data would be to create the training, validation and test sets first and process the data within each set. However, it is important to keep in mind that only parameters computed from the training set can be used for transforming the data in the validation and test sets (Fig. \ref{fig_data_leak_ex_std}). 
In time series data, a temporal cutoff may be useful in preventing leaking any information about the future, to ensure that any data used for training does not include records with a timestamp later than the cutoff value. 
\\
Another common error leading to leakage is data duplication, when the dataset contains identical or near identical data (\cite{kapoor2022leakage}). For example, when working on sequences of pictures from camera traps, duplicates may correspond to images from the same temporal sequence  (Figure \ref{fig_data_leak_duplication}). In this case, data leakage may happen because the training and validation sets contain the same information even though they correspond to different observations, i.e., different pictures from the same temporal sequence.\\

As a general rule, if the model is “too good to be true” (e.g. with over-excellent evaluation metrics), we should get suspicious and check for potential data leakage. Again, we advise using a holdout dataset (i.e., test set) as a final sanity check for model performance and generality.

\section{Treat imbalanced datasets with care}\label{imbalancedData} 
In ecological data, it is common for one or several classes to be more frequent than other(s) (e.g. sporadic distribution vegetation types, apex predators in camera trap images).
Imbalance distribution in data can lead to poor predictive performance in minority classes as conventional classification models tend to be biased towards majority classes (e.g. in deep learning \cite{bud18}).
In those cases, it is challenging for models to learn the characteristics of the observations from the minority class and to differentiate those observations from the others \cite{fernandez2018learning}. 
Indeed, by construction, the rarest classes are under-represented in the loss function and its numerical minimization tend to be driven by the frequent classes. Ignoring data imbalance while building a classification model generally lead to poor predictive performance on the minority class (see Box \ref{lynxbox}).
This is problematic as minority class(es) are often the class(es) of interest (e.g. rare species \cite{wea19}) or rare habitats) and reliable model performance in predicting those instances is therefore particularly critical.

 Often, collecting more data will not solve the issue as minority classes are by nature difficult to sample (rare species, rare habitat, rare event) and data imbalance will persist. In these cases, multiple methods have been developed to handle the imbalanced data problem \cite{he2009learning,bud18}. A common technique is to use resampling approaches (see Box \ref{lynxbox}). These techniques work at the data level by modifying the number of instances in majority and minority classes to balance the data distribution independently of the learning algorithm \cite{leevy2018survey}. With undersampling, the most abundant classes are down-sampled. On the contrary, with oversampling, the rarest classes are over-sampled. In the later case, data augmentation is the leading method (instead of duplication). It consists of generating new data from existing observations by using some disruptions and changes (e.g. changing orientation or colors in images; drawing a small subset of variables from random distributions in ecological studies). 
Another option is to assign different weights in the loss function to the observations belonging to the majority or minority class \cite{chen2004rf}. However, this approach is very empirical since it remains challenging to select the optimal weights. Finally, it is also possible to calibrate the classification scores given by a model: the user can try to rescale the scores of each class to improve the classification performance \cite{goorbergh2022harm}.

\section{Choose evaluation metrics carefully}\label{evalMetric} 
An important step before making any prediction on a sample of unseen data is to make sure the model consistently achieves a desirable performance. Various metrics exist to do so and the choice of the metric(s) to use depends on the type of model considered and the problem to solve. 
Using the wrong metric may lead to select poorly performing models ultimately altering the predictions \cite{ferri2009experimental}. Evaluation metrics can also provide deeper insights into the results as they weight the importance of different characteristics in the predictions (Box \ref{lynxbox}). The confusion matrix, while not an evaluation metric, is also a useful tool that compares the model predictions to the actual classes and provide valuable information about the type of errors the model is making (see Box \ref{explainbox} for an example). 
Imbalanced classification problems also complicate the evaluation of predictive performance as popular classification metrics generally assume a balanced class distribution and may be misleading when data are imbalanced \cite{huang2005using, loyola2016study}. 

Therefore, which metrics should the user choose? Top-k accuracy? Sensitivity/specificity? Precision/recall? There is no general answer here. Depending on the problem, some predictive errors may be more serious than others. For instance, in some applications, it could be more important to reduce the number of false positives to zero, while a trade-off between a small amount of false positives and false negatives could be preferable in other cases. We recommend to move beyond textbook examples and take the time to convert the objectives of the machine learning approach into the appropriate metrics. \\

\section{Look out for shortcut learning}\label{shortcutlearning} 
\begin{figure}[H]
\centering
\includegraphics[scale = 0.35]
{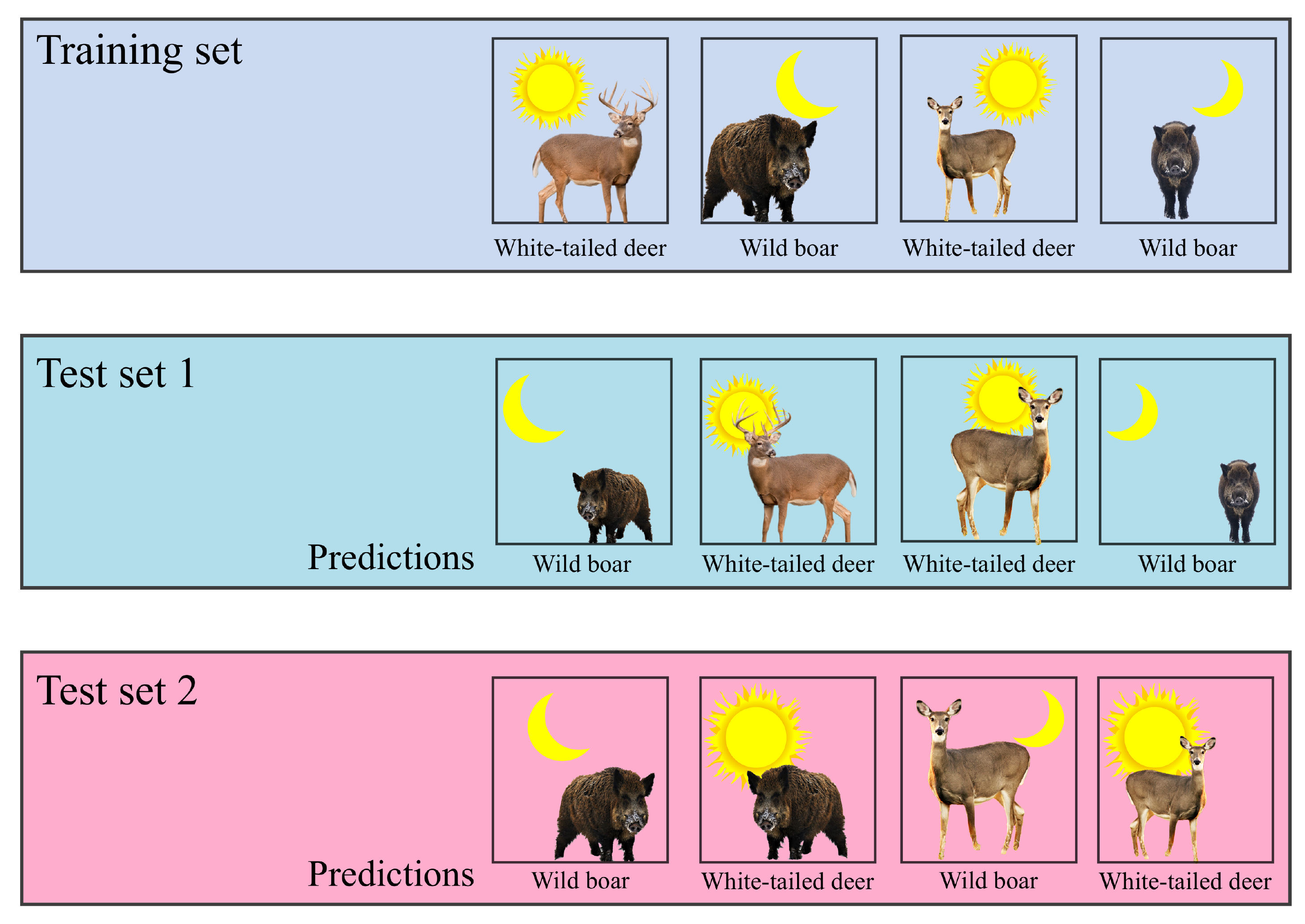}\vspace{0.3cm}
\caption{Example of shortcut learning opportunity in a neural network that aims to classify species in images. During training, white-tailed deer pictures were always taken in daylight; wild boar pictures in the nightime. This pattern is still present in test set 1 (middle row) but not in test set 2 (bottom row), exposing the shortcut: the model has learned to associate the picture timestamp to the species. On test set 2, the predictions are therefore erroneous.}
\label{fig_shortcut_learning}
\end{figure}

Due to the black box nature of some algorithms (e.g., neural networks, see Tip \ref{interpretability}), it is often difficult to understand why those models are successful and, in particular, which part of the data and decision rules they choose to focus on when making predictions \cite{geirhos2020shortcut}. 
Shortcuts is a particular group of decision rules based on unintended correlations and other biases in data that the model uses to make predictions. While superficially successful (i.e. perform well on standard benchmarks), these shortcut strategies typically lack generalisation and cause the model to fail unexpectedly (i.e., make inaccurate predictions) when transferring to slightly different data \cite{geirhos2020shortcut}. In Figure \ref{fig_shortcut_learning}, we show an example of data that could lead to a shortcut opportunity. In this example, the data used by the model to learn how to differentiate two species, the wild boar and the white-tailed deer, included only nocturnal pictures of wild boar and diurnal pictures of deer. 
In that scenario, the model may learn to recognize species by focusing on image timestamps rather than by learning more complex shapes and patterns of the animals themselves. The variable day/night may be used as an unintended predictor for the species identification. 

Several approaches can be used to limit shortcut opportunities. First, many shortcuts are a consequence of natural relationships (e.g., between a species and its typical surrounding landscapes \cite{beery2018recognition, schneider2020three}) and can be avoided by modifying the training data to restrict the model’s access to shortcut features. In our example in Figure \ref{fig_shortcut_learning}, including both nocturnal and diurnal pictures of each species would block the model from learning the shortcut feature day/night as a predictor to recognize the species. Adding noise to the training data with data augmentation  may also be a solution to discourage the model from learning unintended relationships with the output. In photo-identification, a common practice is to crop the picture around the area of interest (e.g., face or individual pattern) in order to promote the training of the model on the zone of interests only  \cite{miele2020revisiting,rigoudy2022deepfaune}. Another recommendation is, again, to use on an out-of-sample test set 
to evaluate the model and test its generalisation beyond the narrowly learned settings \cite{geirhos2020shortcut}.

\section{Add some transparency in your black box models}\label{interpretability} 
Machine learning models lie on a continuum of interpretability and complexity and high predictive performance may come with a loss in interpretability \cite{carmichael2018data, lucas2020translucent, freitas2014comprehensible, du2019techniques}. Highly performing but complex models generally offer less visibility on how predictions are made, how the explanatory input variables impact the output and what the relationships between variables are \cite{lucas2020translucent, chakraborty2017interpretability}. They are therefore often considered to be black box models, hard to understand and communicate to a target audience. On the other side of the continuum, models like regressions or shallow decision trees (i.e., small trees with low depth) are simple and easy to understand when used with a few variables but may not be optimal for prediction. Random forests also offer the possibility to inspect variables importance, to enhance interpretability. Ecologists should be aware of this continuum before choosing which machine learning model to use. A model that is appropriate for a study focusing on prediction is unlikely to be optimal if the aim of the study is to understand the impact and directionality of the relationships between the explanatory input variables and the output. 

However, regardless of the end goal of the study, some level of interpretability remains indispensable to validate and improve models \cite{linardatos2020explainable} and to avoid dangerous traps (e.g., shortcut learning, see Tip \ref{shortcutlearning}). 
Increasing research has been aiming at helping to explain predictions made by complex models \cite{lucas2020translucent} and various methods are now available (e.g., SHAP \cite{lundberg2017unified}, LIME \cite{ribeiro2016model}). If the selected model is not interpretable {\it per se} (e.g. a neural network), there is still a path to gain transparency on the underlying process. For example, in deep learning for images classification, heatmaps have been used to highlight the image zones that were selected/activated by the model \cite{ras2022explainable} (e.g. the curved tail for baboons identification \cite{mia19}). In Box \ref{explainbox}, we show how two other methods could improve the transparency of models usually considered to be black box models.

\section{Make sure you do not learn from errors}\label{Misclassification}
 Real-world data contain redundancy, duplicates and mislabelled classes that can significantly reduce machine learning efficiency \cite{kotsiantis2006data}. This is a particularly known issue in data collected from various citizen science programs \cite{kos16,meschini2021reliability} or in datasets that have been merged from different sources before being fed into a model.
  To avoid a loss in model performance, an important effort must be made to curate, clean and prepare the data before training any model. This task may be time-consuming and hectic but often provides a greater payoff than experimenting with any advanced modelling approaches. 

We suggest to consider machine learning as a tool to facilitate and automate the data cleaning process. In Box \ref{vegebox}, we showed how a cross-validation approach could be used to flag observations in the training set that were likely to be mislabelled. In this example, we only used the model prediction probabilities to identify the observations that required our attention. However, more elaborate tools that implement a family of theory and algorithms called {\it confident learning} have recently been developed \cite{northcutt2019confidentlearning}. They can detect flaws in datasets, characterize label noises, find label errors, fix datasets and improve the model performance by training on cleaned data with just a few lines of code (see open-source package \texttt{cleanlab} \cite{northcutt2019confidentlearning}).

\section*{Conclusion}\label{conclusion}
To face pressing challenges such as climate change and biodiversity loss, precise ecological predictions are critically needed by policy makers and ecosystems managers \cite{clark2001ecological}. Due to their high predictive performance and flexibility, machine learning models are an appropriate and efficient tool for ecologists. 
In this paper, we shared a few tips to help ecologists that are getting started with machine learning to avoid the common mistakes and traps and overcome some of the known challenges of those models. We believe that the use of machine learning by ecologists could result in important advances in ecology. Machine learning approaches also have the potential to be used for more than just model building and prediction, e.g. data cleaning, hypothesis creation and testing and discovery of new patterns in unlabelled data, making these approaches a powerful and valuable tool in the ecologist's toolbox. \\

\section*{Acknowledgments}
This work was supported by a grant from the French National Research Agency (grant ANR-16-CE02-0007). We warmly thank Christophe Duchamp (OFB, France) and Fridolin Zimmermann (KORA, Switzerland) for sharing the data on lynx collisions with vehicles.

\bibliographystyle{unsrt}
\bibliography{main}

\newpage
\input{boxes}


\end{document}

%% file: boxes.tex
\begin{numberedbox}[label={logisticbox}]{Logistic regression in the machine learning mindset}
We tackle here a binary classification problem using a logistic regression. We want to estimate the parameter of the model using a machine learning approach.
\lstset{style=mystyle}
\begin{lstlisting}
## Sepal/Petal length as predictors,
## binary class (being or not Setosa) as response
## coded with 0="non setosa" and 1="setosa"
> data(iris)
> full.set <- data.frame(Sepal.Length=iris$Sepal.Length,
                         Petal.Length=iris$Petal.Length,
                         Setosa=as.integer(iris$Species=="setosa"))

Sepal.Length Petal.Length Setosa
1          5.1          1.4      1
2          4.9          1.4      1
...
149        6.2          5.4      0
150        5.9          5.1      0

## Building a train and a validation dataset
## with a random split (for simplicity; see drawbacks in Tip 2)
> idx <- sample(1:150)
> train.set <- full.set[idx[1:120],]
> val.set <- full.set[idx[121:150],]

## Loss function measuring the gap between true class and prediction 
> loss <- function(par){
    a <- par[1]
    b <- par[2]
    c <- par[3]
    proba <- 1/(1+exp(-(a + b*train.set$Petal.Length+ c*train.set$Sepal.Length)))
    loss <- sum(-train.set$Setosa*log(proba) - (1-train.set$Setosa)*log(1-proba)) / length(train.set$Setosa)
    print(loss)
    return(loss)
}
   
## Training the model by minimizing the loss
## to find the optimal parameters using train data set
> bestpar <- optim(
    par = c(a = 0, b = 0, d=0),
    fn = loss
)$par

[1] 0.8425373
[1] 0.8421624
[1] 0.6474087
...
[1] 4.880504e-08

> cat("Best params are:", bestpar,"\n")

Best params are: 2.407035 -23.63244 10.78046 

> cat("Minimal train loss is:", loss(bestpar),"\n")

Best train loss is: 4.742235e-08 

## Building a predictor
## that predicts a binary class (being or not Setosa)
> predict <- function(data.set){
    a = bestpar[1]
    b = bestpar[2]
    c = bestpar[3]
    proba <- 1/(1+exp(-(a + b*data.set$Petal.Length+ c*data.set$Sepal.Length)))
    return(as.integer(proba>0.5))
}

## Checking prediction on validation data set
> head(data.frame(prediction=predict(val.set), truth=val.set$Setosa))

   prediction truth
1           0     0
2           1     1
3           1     1
...

## Predicting on a new test data set
> test.set <- data.frame(Petal.Length=c(2,3.5),
                       Sepal.Length=c(5.5,6),
                       Setosa=c(NA,NA))

  Petal.Length Sepal.Length Setosa
1          2.0          5.5     NA
2          3.5          6.0     NA

> test.set$Setosa <- predict(test.set)

  Petal.Length Sepal.Length Setosa
1          2.0          5.5      1
2          3.5          6.0      0
\end{lstlisting}
\end{numberedbox}

\begin{numberedbox}[label={lynxbox}]{Unbalanced data and metrics to estimate risk of collision with vehicles in Lynx}

We used animal-vehicle collision data collected on the Eurasian Lynx in the French Jura Mountains to predict animals at high risk of collision in the Swiss Jura Mountains. 
Data on collisions were collected from 1982 to 2018, in France by OFB (the French Biodiversity Agency  \url{https://www.ofb.gouv.fr/en}) and in Switzerland by KORA Carnivore Ecology and Wildlife Management (\url{https://www.kora.ch/en/}). A grid of 1 km$^2$ cells was overlaid on the Swiss-French road network taken from Open Street Map. Explanatory variables were urban land use cover (from the Corine Land Cover 2012 data base \url{https://land.copernicus.eu/pan-european/corine-land-cover/clc-2012}), distance to major road segments, human density, road class (proximal measure of traffic intensity, split into highways, main, local and regional roads), and total length of road segments. We also used lynx presence that we summarised over the study period by the cumulated number of times a cell was occupied (data were provided by the French Biodiversity Agency \url{https://carmen.carmencarto.fr/38/Lynx.map}).\\

\begin{table}[H]
\centering
\begin{tabular}{ c c c }
 Origin dataset & No Collisions & Collisions \\ \hline\hline
 France & 11238 & 80 \\ \hline
 Switzerland & 9472 & 69 \\ \hline
\end{tabular}\vspace{0.3cm}
\captionsetup{justification=centering}
\caption{Number of individuals in each class for French and Swiss datasets.}
\label{lynx_imbalance}
\end{table}

This dataset is characterised by a strong imbalance in the response variable (Table \ref{lynx_imbalance}). We modelled the data using random forest. Our results showed that the classifier not accounting for the imbalance in data (first row in Table\ref{table_confusion_matrix})  was unable to recognize any instance from the minority class (i.e. high risk individuals). 
We therefore used resampling methods: 1) undersampling, 2) oversampling, 3) combined under and oversampling and 4) SMOTE \cite{chawla2002smote}. 
However, implementing resampling methods in the random forest barely improved the model predictive power in the minority class (Table \ref{table_confusion_matrix} and Fig. \ref{fig_evaluation_metrics}).\\

\begin{table}[H]
\centering
\begin{tabular}{|c||c|c|c|c|}
\hline
Model & TP & TN & FP & FN \\ \hline\hline
No sampling   & 0  & 9472 & 0    & 69 \\ \hline
Oversampling  & 1  & 9469 & 3    & 68 \\ \hline
Undersampling & 63 & 3198 & 6274 & 6  \\ \hline
Combined      & 1  & 9466 & 6    & 68 \\ \hline
SMOTE         & 1  & 9468 & 4    & 68 \\ \hline
    \end{tabular}\vspace{0.3cm}
\caption{Predictive performance computed from the confusion matrix obtained for each model. \textit{Abbreviations}: TP: True Positive; TN: True Negative; FP: False Positive; FN: False Negative}
\label{table_confusion_matrix}
\end{table}

This study also highlighted the importance of choosing an appropriate evaluation metric as the classification accuracy suggested outstanding model performance in most models despite their inability to predict the class of interest. While the undersampling model predicted significantly more positive instances (i.e., higher recall), the number of false positives increased substantially (i.e. very low precision, low F-scores). Therefore, none of those models seemed to be informative for the end-user (i.e., low F-Score for all models). To improve the model predictive performance and handle the imbalance in data, other options should be considered in this particular case (see tip \ref{imbalancedData}).\\

\begin{figure}[H]
\centering
\includegraphics[width=\textwidth]
{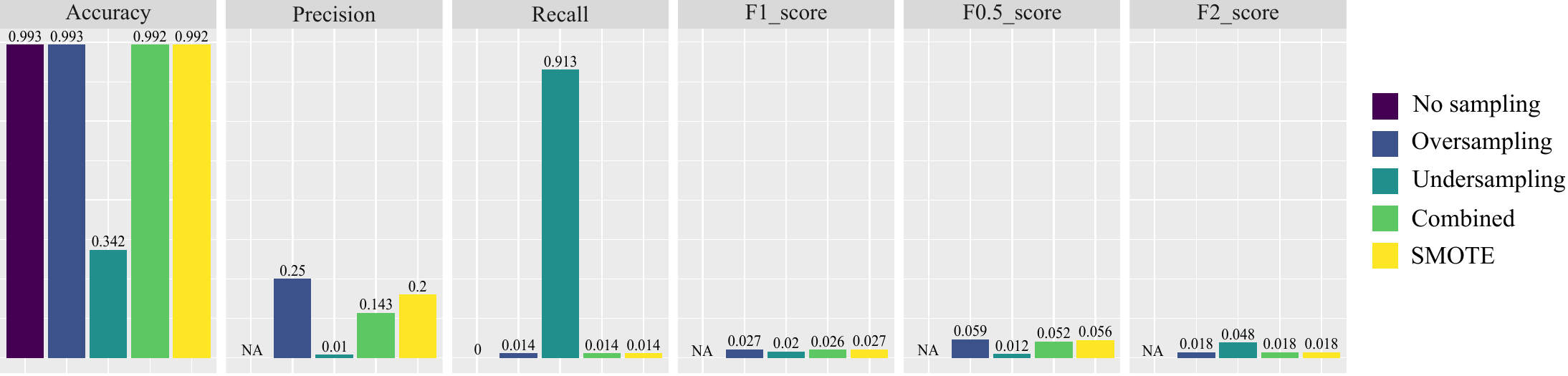}\vspace{0.3cm}
\caption{Accuracy, precision, recall and F-scores for the model without and with resampling methods}
\label{fig_evaluation_metrics}
\end{figure}

R scripts to reproduce this study are provided in Supplementary Material.
\end{numberedbox}

\begin{numberedbox}[label={explainbox}]{Explaining predictions in species distribution modeling.}
We used the dataset available in the supplementary material of Zurell et al. \cite{zurell2020testing} that recorded information about the presence or absence of the Ring Ouzel associated with 52 environmental predictors and investigated the drivers of the predictions in each model.
We used two different machine learning models, a random forest and an artificial neural network, to predict the presence of the Ring Ouzel (\textit{Turdus torquatus}) in Switzerland. 
Both model predicted the presence or absence of the Ring Ouzel with a high accuracy, 0.90 for the neural network and 0.92 for the random forest respectively.\\
%

 To investigate which variables played an important role in those predictions, we generated the features importance for each model. In random forests, features importance are directly provided during the training and validation step. The variable ranks are based on the Gini importance score \cite{breiman2017classification} (Fig. \ref{fig_gini}) or permutation importance measure \cite{breiman2001random} (not shown).

\begin{figure}[H]
\centering
\includegraphics[scale = 0.35]
{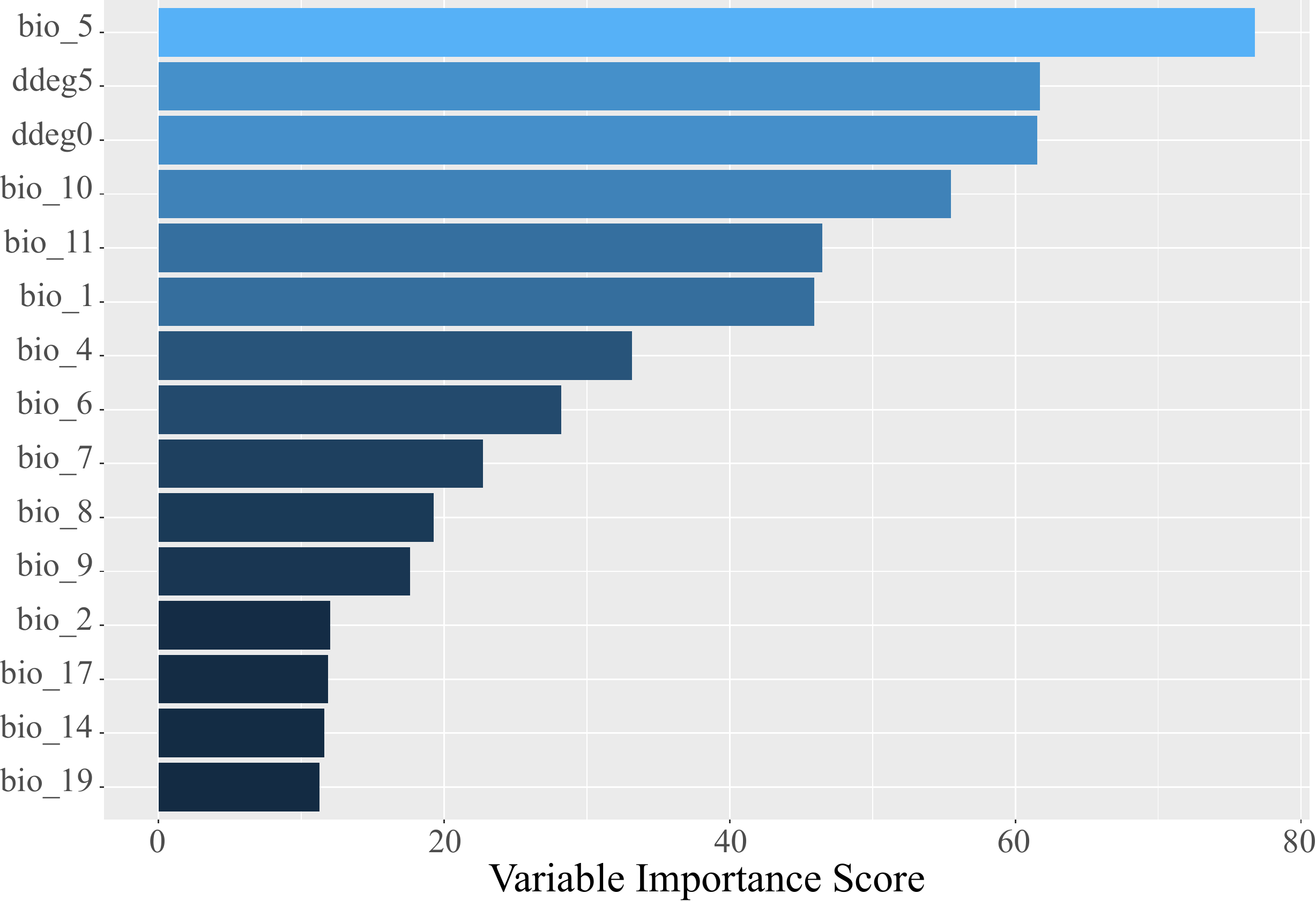}\vspace{0.3cm}
\caption{Variable importance according to the Gini importance measure generated by the random forest model}
\label{fig_gini}
\end{figure}


With neural networks, generating feature importance is not as straightforward. Feature importance can be determined by calculating the permutation importance but the implementation needs to be done by the user himself.  In our example, we used another technique, the LIME approach \cite{ribeiro2016model}, to provide a local model interpretability instead of interpretability from the perspective of the entire dataset. The output of LIME explains the contribution of each feature to the prediction of a data sample (Fig. \ref{fig_lime_ann}). For example, in our study, low values of the variable ‘ddeg5’ correlated with the presence (positive cases in Fig. \ref{fig_lime_ann}) of Ring Ouzel. \\

\begin{figure}[H]
\centering
\includegraphics[scale = 0.35]
{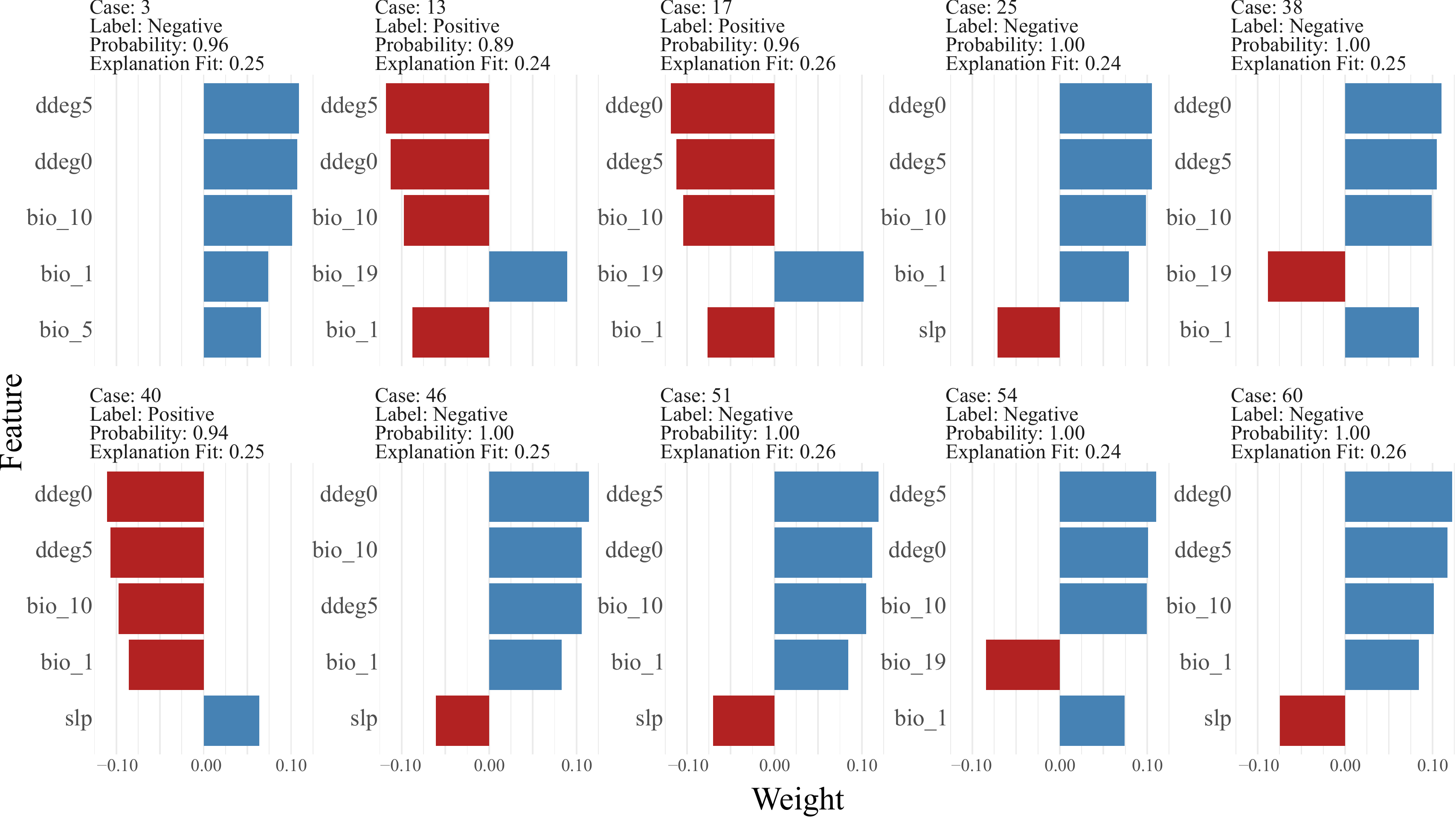}\vspace{0.3cm}
\caption{Visualization of the 5 most important variables driving the predictions in the neural network model for 10 data points (the case number) using the LIME method . Features that have positive correlations with the output are shown in blue, negatively correlated features are shown in red.}
\label{fig_lime_ann}
\end{figure}

R scripts to reproduce this study are provided in Supplementary Material.
\end{numberedbox}

\begin{numberedbox}[label={vegebox}]{Flagging label errors in vegetation surveys}
We used a simulated set of vegetation surveys reporting the abundance of 100 species for 300 locations. Each survey was simulated accorded to three vegetation classes, using three different species assembly schemes that we called A, B and C (response variable). Surveys 1-100 were simulated with type A, 101-250 with type B and 251-300 with type C. \\

The classification information for three surveys (1, 101 and 251) was intentionally modified to introduce labelling errors: survey 1 was labelled as belonging to class B instead of true class A, survey 101 to class C instead of B, and survey 251 to class A instead of C.\\

We trained an XGBoost model \cite{chen2016xgboost} to classify surveys in classes A,B or C. We observed the predictive probabilities obtained from the XGBoost model on the training set (see Fig. \ref{fig_labelling_errors}) using a 3-fold cross-validation approach. 
Survey 251 was labelled as A but the predictive probabilities obtained from the model were close to 0.0 for that particular class. However, the probability was close to 1.0 for class C (Fig. \ref{fig_labelling_errors}). This suggested that survey 251 was wrongly labelled as class A and actually belonged to class C. Similarly, the model successfully flagged the class of surveys 1 and 101 as labelling errors (survey 1 being labelled B while actually belonging to class A and survey 101 classified in class C while being an instance from class C). 

\begin{figure}[H]
\centering
\includegraphics[scale = 1.15]
{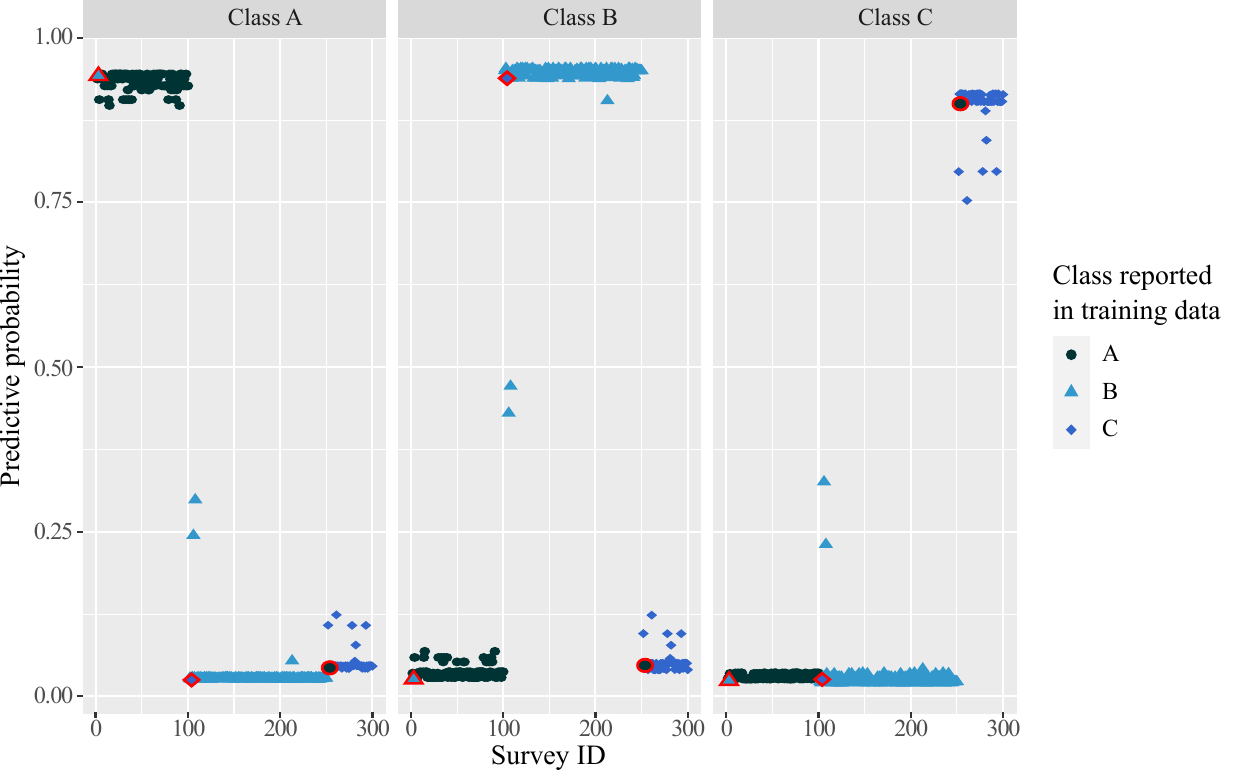}\vspace{0.3cm}
\caption{XGBoost predictive probabilities obtained for each observation and vegetation class. Labels reported in the training data for each observation (i.e., vegetation classes A, B and C) are represented by different symbols and colors (see legend on graph). The three mislabeled training observations were highlighted in red.
}
\label{fig_labelling_errors}
\end{figure}

R scripts to reproduce this study are provided in Supplementary Material.
\end{numberedbox}

%% file: main.bbl
\begin{thebibliography}{10}

\bibitem{cutler2007random}
D~Richard Cutler, Thomas~C Edwards~Jr, Karen~H Beard, Adele Cutler, Kyle~T
  Hess, Jacob Gibson, and Joshua~J Lawler.
\newblock Random forests for classification in ecology.
\newblock {\em Ecology}, 88(11):2783--2792, 2007.

\bibitem{roberts2017cross}
David~R Roberts, Volker Bahn, Simone Ciuti, Mark~S Boyce, Jane Elith, Gurutzeta
  Guillera-Arroita, Severin Hauenstein, Jos{\'e}~J Lahoz-Monfort, Boris
  Schr{\"o}der, Wilfried Thuiller, et~al.
\newblock Cross-validation strategies for data with temporal, spatial,
  hierarchical, or phylogenetic structure.
\newblock {\em Ecography}, 40(8):913--929, 2017.

\bibitem{christin2019applications}
Sylvain Christin, {\'E}ric Hervet, and Nicolas Lecomte.
\newblock Applications for deep learning in ecology.
\newblock {\em Methods in Ecology and Evolution}, 10(10):1632--1644, 2019.

\bibitem{rovero2013camera}
Francesco Rovero, Fridolin Zimmermann, Duccio Berzi, and Paul Meek.
\newblock " which camera trap type and how many do i need?" a review of camera
  features and study designs for a range of wildlife research applications.
\newblock {\em Hystrix}, 2013.

\bibitem{schneider2020three}
Stefan Schneider, Saul Greenberg, Graham~W Taylor, and Stefan~C Kremer.
\newblock Three critical factors affecting automated image species recognition
  performance for camera traps.
\newblock {\em Ecology and evolution}, 10(7):3503--3517, 2020.

\bibitem{geirhos2020shortcut}
Robert Geirhos, J{\"o}rn-Henrik Jacobsen, Claudio Michaelis, Richard Zemel,
  Wieland Brendel, Matthias Bethge, and Felix~A Wichmann.
\newblock Shortcut learning in deep neural networks.
\newblock {\em Nature Machine Intelligence}, 2(11):665--673, 2020.

\bibitem{humphries2018machine}
Grant~RW Humphries, Dawn~R Magness, Falk Huettmann, et~al.
\newblock {\em Machine learning for ecology and sustainable natural resource
  management}.
\newblock Springer, 2018.

\bibitem{pichler2022machine}
Maximilian Pichler and Florian Hartig.
\newblock Machine learning and deep learning--a review for ecologists.
\newblock {\em Methods in Ecology and Evolution}, 2023.

\bibitem{waldchen2018machine}
Jana W{\"a}ldchen and Patrick M{\"a}der.
\newblock Machine learning for image based species identification.
\newblock {\em Methods in Ecology and Evolution}, 9(11):2216--2225, 2018.

\bibitem{valletta2017applications}
John~Joseph Valletta, Colin Torney, Michael Kings, Alex Thornton, and Joah
  Madden.
\newblock Applications of machine learning in animal behaviour studies.
\newblock {\em Animal Behaviour}, 124:203--220, 2017.

\bibitem{gobeyn2019evolutionary}
Sacha Gobeyn, Ans~M Mouton, Anna~F Cord, Andrea Kaim, Martin Volk, and Peter~LM
  Goethals.
\newblock Evolutionary algorithms for species distribution modelling: A review
  in the context of machine learning.
\newblock {\em Ecological Modelling}, 392:179--195, 2019.

\bibitem{tuia2022perspectives}
Devis Tuia, Benjamin Kellenberger, Sara Beery, Blair~R Costelloe, Silvia Zuffi,
  Benjamin Risse, Alexander Mathis, Mackenzie~W Mathis, Frank van Langevelde,
  Tilo Burghardt, et~al.
\newblock Perspectives in machine learning for wildlife conservation.
\newblock {\em Nature communications}, 13(1):1--15, 2022.

\bibitem{dugan2015dcl}
Peter~J Dugan, Christopher~W Clark, Yann~A LeCun, and Sofie~M Van~Parijs.
\newblock Dcl system using deep learning approaches for land-based or
  ship-based real time recognition and localization of marine mammals.
\newblock Technical report, Bioacoustics Research Program, Cornell University
  Ithaca United States, 2015.

\bibitem{shamir2014classification}
Lior Shamir, Carol Yerby, Robert Simpson, Alexander~M von Benda-Beckmann, Peter
  Tyack, Filipa Samarra, Patrick Miller, and John Wallin.
\newblock Classification of large acoustic datasets using machine learning and
  crowdsourcing: Application to whale calls.
\newblock {\em The Journal of the Acoustical Society of America},
  135(2):953--962, 2014.

\bibitem{acevedo2009automated}
Miguel~A Acevedo, Carlos~J Corrada-Bravo, H{\'e}ctor Corrada-Bravo, Luis~J
  Villanueva-Rivera, and T~Mitchell Aide.
\newblock Automated classification of bird and amphibian calls using machine
  learning: A comparison of methods.
\newblock {\em Ecological Informatics}, 4(4):206--214, 2009.

\bibitem{czernecki2018machine}
Bartosz Czernecki, Jakub Nowosad, and Katarzyna Jab{\l}o{\'n}ska.
\newblock Machine learning modeling of plant phenology based on coupling
  satellite and gridded meteorological dataset.
\newblock {\em International journal of biometeorology}, 62(7):1297--1309,
  2018.

\bibitem{xin2020evaluations}
Qinchuan Xin, Jing Li, Ziming Li, Yaoming Li, and Xuewen Zhou.
\newblock Evaluations and comparisons of rule-based and machine-learning-based
  methods to retrieve satellite-based vegetation phenology using modis and usa
  national phenology network data.
\newblock {\em International Journal of Applied Earth Observation and
  Geoinformation}, 93:102189, 2020.

\bibitem{norouzzadeh2018automatically}
Mohammad~Sadegh Norouzzadeh, Anh Nguyen, Margaret Kosmala, Alexandra Swanson,
  Meredith~S Palmer, Craig Packer, and Jeff Clune.
\newblock Automatically identifying, counting, and describing wild animals in
  camera-trap images with deep learning.
\newblock {\em Proceedings of the National Academy of Sciences},
  115(25):E5716--E5725, 2018.

\bibitem{mohanty2016using}
Sharada~P Mohanty, David~P Hughes, and Marcel Salath{\'e}.
\newblock Using deep learning for image-based plant disease detection.
\newblock {\em Frontiers in plant science}, 7:1419, 2016.

\bibitem{van08}
Laurens Van~der Maaten and Geoffrey Hinton.
\newblock Visualizing data using t-sne.
\newblock {\em Journal of machine learning research}, 9(11), 2008.

\bibitem{mci18}
Leland McInnes, John Healy, and James Melville.
\newblock Umap: Uniform manifold approximation and projection for dimension
  reduction.
\newblock {\em arXiv preprint arXiv:1802.03426}, 2018.

\bibitem{set20}
Sarab~S Sethi, Nick~S Jones, Ben~D Fulcher, Lorenzo Picinali, Dena~Jane Clink,
  Holger Klinck, C~David~L Orme, Peter~H Wrege, and Robert~M Ewers.
\newblock Characterizing soundscapes across diverse ecosystems using a
  universal acoustic feature set.
\newblock {\em Proceedings of the National Academy of Sciences},
  117(29):17049--17055, 2020.

\bibitem{gareth2013introduction}
James Gareth, Witten Daniela, Hastie Trevor, and Tibshirani Robert.
\newblock {\em An introduction to statistical learning: with applications in
  R}.
\newblock Spinger, 2013.

\bibitem{chr21}
Sylvain Christin, {\'E}ric Hervet, and Nicolas Lecomte.
\newblock Going further with model verification and deep learning.
\newblock {\em Methods in Ecology and Evolution}, 12(1):130--134, 2021.

\bibitem{ying2019overview}
Xue Ying.
\newblock An overview of overfitting and its solutions.
\newblock In {\em Journal of physics: Conference series}, volume 1168, page
  022022. IOP Publishing, 2019.

\bibitem{dietterich1995overfitting}
Tom Dietterich.
\newblock Overfitting and undercomputing in machine learning.
\newblock {\em ACM computing surveys (CSUR)}, 27(3):326--327, 1995.

\bibitem{whytock2021robust}
Robin~C Whytock, Jedrzej {\'S}wie{\.z}ewski, Joeri~A Zwerts, Tadeusz
  Bara-S{\l}upski, Aur{\'e}lie~Flore Koumba~Pambo, Marek Rogala, Laila Bahaa-el
  din, Kelly Boekee, Stephanie Brittain, Anabelle~W Cardoso, et~al.
\newblock Robust ecological analysis of camera trap data labelled by a machine
  learning model.
\newblock {\em Methods in Ecology and Evolution}, 12(6):1080--1092, 2021.

\bibitem{kuhn2013applied}
Max Kuhn, Kjell Johnson, et~al.
\newblock {\em Applied predictive modeling}, volume~26.
\newblock Springer, 2013.

\bibitem{kapoor2022leakage}
Sayash Kapoor and Arvind Narayanan.
\newblock Leakage and the reproducibility crisis in ml-based science.
\newblock {\em arXiv preprint arXiv:2207.07048}, 2022.

\bibitem{kouw2018introduction}
Wouter~M Kouw and Marco Loog.
\newblock An introduction to domain adaptation and transfer learning.
\newblock {\em arXiv preprint arXiv:1812.11806}, 2018.

\bibitem{machireddy2022continual}
Amrutha Machireddy, Ranganath Krishnan, Nilesh Ahuja, and Omesh Tickoo.
\newblock Continual active adaptation to evolving distributional shifts.
\newblock In {\em Proceedings of the IEEE/CVF Conference on Computer Vision and
  Pattern Recognition}, pages 3444--3450, 2022.

\bibitem{lecun2015deep}
Yann LeCun, Yoshua Bengio, and Geoffrey Hinton.
\newblock Deep learning.
\newblock {\em nature}, 521(7553):436--444, 2015.

\bibitem{rigoudy2022deepfaune}
Noa Rigoudy, Abdelbaki Benyoub, Aurelien Besnard, Carole Birck, Yoann Bollet,
  Yoann Bunz, Nina De~Backer, Gerard Caussimont, Anne Delestrade, Lucie Dispan,
  et~al.
\newblock The deepfaune initiative: a collaborative effort towards the
  automatic identification of the french fauna in camera-trap images.
\newblock {\em bioRxiv}, 2022.

\bibitem{volk2019towards}
Georg Volk, Stefan M{\"u}ller, Alexander Von~Bernuth, Dennis Hospach, and
  Oliver Bringmann.
\newblock Towards robust cnn-based object detection through augmentation with
  synthetic rain variations.
\newblock In {\em 2019 IEEE Intelligent Transportation Systems Conference
  (ITSC)}, pages 285--292. IEEE, 2019.

\bibitem{mumuni2022data}
Alhassan Mumuni and Fuseini Mumuni.
\newblock Data augmentation: A comprehensive survey of modern approaches.
\newblock {\em Array}, page 100258, 2022.

\bibitem{uhlmann2019scientific}
Eric~Luis Uhlmann, Charles~R Ebersole, Christopher~R Chartier, Timothy~M
  Errington, Mallory~C Kidwell, Calvin~K Lai, Randy~J McCarthy, Amy Riegelman,
  Raphael Silberzahn, and Brian~A Nosek.
\newblock Scientific utopia iii: Crowdsourcing science.
\newblock {\em Perspectives on Psychological Science}, 14(5):711--733, 2019.

\bibitem{tod23}
Lindsay~C Todman, Alex Bush, and Amelia~SC Hood.
\newblock ‘small data’for big insights in ecology.
\newblock {\em Trends in Ecology \& Evolution}, 2023.

\bibitem{pan21}
Omiros Pantazis, Gabriel~J Brostow, Kate~E Jones, and Oisin Mac~Aodha.
\newblock Focus on the positives: Self-supervised learning for biodiversity
  monitoring.
\newblock In {\em Proceedings of the IEEE/CVF International Conference on
  Computer Vision}, pages 10583--10592, 2021.

\bibitem{kaufman2012leakage}
Shachar Kaufman, Saharon Rosset, Claudia Perlich, and Ori Stitelman.
\newblock Leakage in data mining: Formulation, detection, and avoidance.
\newblock {\em ACM Transactions on Knowledge Discovery from Data (TKDD)},
  6(4):1--21, 2012.

\bibitem{bud18}
Mateusz Buda, Atsuto Maki, and Maciej~A Mazurowski.
\newblock A systematic study of the class imbalance problem in convolutional
  neural networks.
\newblock {\em Neural networks}, 106:249--259, 2018.

\bibitem{fernandez2018learning}
Alberto Fern{\'a}ndez, Salvador Garc{\'\i}a, Mikel Galar, Ronaldo~C Prati,
  Bartosz Krawczyk, and Francisco Herrera.
\newblock {\em Learning from imbalanced data sets}, volume~10.
\newblock Springer, 2018.

\bibitem{wea19}
Oliver~R Wearn, Robin Freeman, and David~MP Jacoby.
\newblock Responsible ai for conservation.
\newblock {\em Nature Machine Intelligence}, 1(2):72--73, 2019.

\bibitem{he2009learning}
Haibo He and Edwardo~A Garcia.
\newblock Learning from imbalanced data.
\newblock {\em IEEE Transactions on knowledge and data engineering},
  21(9):1263--1284, 2009.

\bibitem{leevy2018survey}
Joffrey~L Leevy, Taghi~M Khoshgoftaar, Richard~A Bauder, and Naeem Seliya.
\newblock A survey on addressing high-class imbalance in big data.
\newblock {\em Journal of Big Data}, 5(1):1--30, 2018.

\bibitem{chen2004rf}
Chao Chen, Andy Liaw, Leo Breiman, et~al.
\newblock Using random forest to learn imbalanced data.
\newblock {\em University of California, Berkeley}, 110(1-12):24, 2004.

\bibitem{goorbergh2022harm}
Ruben van~den Goorbergh, Maarten van Smeden, Dirk Timmerman, and Ben
  Van~Calster.
\newblock The harm of class imbalance corrections for risk prediction models:
  illustration and simulation using logistic regression.
\newblock {\em arXiv preprint arXiv:2202.09101}, 2022.

\bibitem{ferri2009experimental}
C{\'e}sar Ferri, Jos{\'e} Hern{\'a}ndez-Orallo, and R~Modroiu.
\newblock An experimental comparison of performance measures for
  classification.
\newblock {\em Pattern recognition letters}, 30(1):27--38, 2009.

\bibitem{huang2005using}
Jin Huang and Charles~X Ling.
\newblock Using auc and accuracy in evaluating learning algorithms.
\newblock {\em IEEE Transactions on knowledge and Data Engineering},
  17(3):299--310, 2005.

\bibitem{loyola2016study}
Octavio Loyola-Gonz{\'a}lez, Jos{\'e}~Fco Mart{\'\i}nez-Trinidad,
  Jes{\'u}s~Ariel Carrasco-Ochoa, and Milton Garc{\'\i}a-Borroto.
\newblock Study of the impact of resampling methods for contrast pattern based
  classifiers in imbalanced databases.
\newblock {\em Neurocomputing}, 175:935--947, 2016.

\bibitem{beery2018recognition}
Sara Beery, Grant Van~Horn, and Pietro Perona.
\newblock Recognition in terra incognita.
\newblock In {\em Proceedings of the European conference on computer vision
  (ECCV)}, pages 456--473, 2018.

\bibitem{miele2020revisiting}
Vincent Miele, Gaspard Dussert, Bruno Spataro, Simon Chamaillé-Jammes,
  Dominique Allainé, and Christophe Bonenfant.
\newblock Revisiting animal photo-identification using deep metric learning and
  network analysis.
\newblock {\em Methods in Ecology and Evolution}, 12(5):863--873, 2021.

\bibitem{carmichael2018data}
Iain Carmichael and JS~Marron.
\newblock Data science vs. statistics: two cultures?
\newblock {\em Japanese Journal of Statistics and Data Science}, 1(1):117--138,
  2018.

\bibitem{lucas2020translucent}
Tim~CD Lucas.
\newblock A translucent box: interpretable machine learning in ecology.
\newblock {\em Ecological Monographs}, 90(4):e01422, 2020.

\bibitem{freitas2014comprehensible}
Alex~A Freitas.
\newblock Comprehensible classification models: a position paper.
\newblock {\em ACM SIGKDD explorations newsletter}, 15(1):1--10, 2014.

\bibitem{du2019techniques}
Mengnan Du, Ninghao Liu, and Xia Hu.
\newblock Techniques for interpretable machine learning.
\newblock {\em Communications of the ACM}, 63(1):68--77, 2019.

\bibitem{chakraborty2017interpretability}
Supriyo Chakraborty, Richard Tomsett, Ramya Raghavendra, Daniel Harborne,
  Moustafa Alzantot, Federico Cerutti, Mani Srivastava, Alun Preece, Simon
  Julier, Raghuveer~M Rao, et~al.
\newblock Interpretability of deep learning models: A survey of results.
\newblock In {\em 2017 IEEE smartworld, ubiquitous intelligence \& computing,
  advanced \& trusted computed, scalable computing \& communications, cloud \&
  big data computing, Internet of people and smart city innovation
  (smartworld/SCALCOM/UIC/ATC/CBDcom/IOP/SCI)}, pages 1--6. IEEE, 2017.

\bibitem{linardatos2020explainable}
Pantelis Linardatos, Vasilis Papastefanopoulos, and Sotiris Kotsiantis.
\newblock Explainable ai: A review of machine learning interpretability
  methods.
\newblock {\em Entropy}, 23(1):18, 2020.

\bibitem{lundberg2017unified}
Scott~M Lundberg and Su-In Lee.
\newblock A unified approach to interpreting model predictions.
\newblock {\em Advances in neural information processing systems}, 30, 2017.

\bibitem{ribeiro2016model}
Marco~Tulio Ribeiro, Sameer Singh, and Carlos Guestrin.
\newblock Model-agnostic interpretability of machine learning.
\newblock {\em arXiv preprint arXiv:1606.05386}, 2016.

\bibitem{ras2022explainable}
Gabrielle Ras, Ning Xie, Marcel van Gerven, and Derek Doran.
\newblock Explainable deep learning: A field guide for the uninitiated.
\newblock {\em Journal of Artificial Intelligence Research}, 73:329--397, 2022.

\bibitem{mia19}
Zhongqi Miao, Kaitlyn~M Gaynor, Jiayun Wang, Ziwei Liu, Oliver Muellerklein,
  Mohammad~Sadegh Norouzzadeh, Alex McInturff, Rauri~CK Bowie, Ran Nathan,
  Stella~X Yu, et~al.
\newblock Insights and approaches using deep learning to classify wildlife.
\newblock {\em Scientific reports}, 9(1):8137, 2019.

\bibitem{kotsiantis2006data}
Sotiris~B Kotsiantis, Dimitris Kanellopoulos, and Panagiotis~E Pintelas.
\newblock Data preprocessing for supervised leaning.
\newblock {\em International journal of computer science}, 1(2):111--117, 2006.

\bibitem{kos16}
Margaret Kosmala, Andrea Wiggins, Alexandra Swanson, and Brooke Simmons.
\newblock Assessing data quality in citizen science.
\newblock {\em Frontiers in Ecology and the Environment}, 14(10):551--560,
  2016.

\bibitem{meschini2021reliability}
Marta Meschini, Mariana Machado~Toffolo, Chiara Marchini, Erik Caroselli,
  Fiorella Prada, Arianna Mancuso, Silvia Franzellitti, Laura Locci, Marco
  Davoli, Michele Trittoni, et~al.
\newblock Reliability of data collected by volunteers: A nine-year citizen
  science study in the red sea.
\newblock {\em Frontiers in Ecology and Evolution}, page 395, 2021.

\bibitem{northcutt2019confidentlearning}
Curtis~G. Northcutt, Lu~Jiang, and Isaac~L. Chuang.
\newblock Confident learning: Estimating uncertainty in dataset labels, 2019.

\bibitem{clark2001ecological}
James~S Clark, Steven~R Carpenter, Mary Barber, Scott Collins, Andy Dobson,
  Jonathan~A Foley, David~M Lodge, Mercedes Pascual, Roger Pielke~Jr, William
  Pizer, et~al.
\newblock Ecological forecasts: an emerging imperative.
\newblock {\em science}, 293(5530):657--660, 2001.

\bibitem{chawla2002smote}
Nitesh~V Chawla, Kevin~W Bowyer, Lawrence~O Hall, and W~Philip Kegelmeyer.
\newblock Smote: synthetic minority over-sampling technique.
\newblock {\em Journal of artificial intelligence research}, 16:321--357, 2002.

\bibitem{zurell2020testing}
Damaris Zurell, Niklaus~E Zimmermann, Helge Gross, Andri Baltensweiler, Thomas
  Sattler, and Rafael~O W{\"u}est.
\newblock Testing species assemblage predictions from stacked and joint species
  distribution models.
\newblock {\em Journal of Biogeography}, 47(1):101--113, 2020.

\bibitem{breiman2017classification}
Leo Breiman, Jerome~H Friedman, Richard~A Olshen, and Charles~J Stone.
\newblock {\em Classification and regression trees}.
\newblock Routledge, 2017.

\bibitem{breiman2001random}
Leo Breiman.
\newblock Random forests.
\newblock {\em Machine learning}, 45(1):5--32, 2001.

\bibitem{chen2016xgboost}
Tianqi Chen and Carlos Guestrin.
\newblock Xgboost: A scalable tree boosting system.
\newblock In {\em Proceedings of the 22nd acm sigkdd international conference
  on knowledge discovery and data mining}, pages 785--794, 2016.

\end{thebibliography}
